\documentstyle[fleqn,12pt]{article}

\begin{document}

\baselineskip=0.74cm
\title{\bf              Degeneracy of Landau levels and quantum group $
sl_{q}(2) $}
\author{               Guang-Hong $\mbox{Chen}^{\dag}$,
                       Le-Man $\mbox{Kuang}^{\dag \ddag}$ \\
                       and Mo-Lin  $\mbox{Ge}^{\dag}$ \\
{\normalsize \it \dag  Theoretical Physics Division,
                       Nankai Institute of Mathematics,}\\
{\normalsize \it       Tianjin 300071, P.R.China}\\
{\normalsize \it \ddag Department of Physics,
                       Hunan Normal University,}\\
{\normalsize \it       Hunan 410006, P.R.China}}

\maketitle

\begin{abstract}
\baselineskip=0.90cm
We show that there is a kind of quantum group symmetry $ sl_{q}(2) $
in the usual Landau problem and it is  this quantum group symmetry
that governs the degeneracy of Landau levels. We find that under the
periodic boundary condition, the degree of degeneracy of Landau levels
is finite, and it just equals  the dimension of the irreducible cyclic
representation of the quantum group $ sl_{q}(2) $.

\end{abstract}
\hspace{0.5cm}
\newpage
 It is well known that the symmetry play an important role in  phyisics.
  Sometimes we need not solve the problem explicitly, we  can obtain  much
important and
interesting information of a physical system or simplify our calculation by
analyzing the symmetry of
  system. In particular, the degeneracy of energy levels is often related
  to some dynamical symmetries of a system [1,2]. In general, for a system
   with Hamitanian $ H $, let  $ \hat{F} $ and $ \hat{G} $ denote  two
    operators corresponding to physical quantities of the system, if they do
not commute
     with eath other, i.e., $ [\hat{F},\hat{G}]\ne 0 $, and both are the
     conservative quantities, then  energy levels
      of the system must be degenerate except for few special levels. For
       example, for a spinless particle moving in a plane, this system has
       the Euclidian group   symmetry whose generators are momentum operator
     $\hat{p}_{i} (i=$x$,$y$)$ and the $z$-component of the angular momentum
 $\hat{L}_{z}$. Since $\hat{p}_{i} (i=$x$,$y$)$ and  $\hat{L}_{z}$ do not
commute
 with each other, and  both of them are the conservative quantities of the
system,
 energy levels of this system exhibit infinite-fold degeneracy which
comes from the infinity of the dimension of the irreducible representation of
the
Euclidian group [2].

On the other hand, in the past ten years, the so-called quantum group symmetry
(QGS) and its representation theory has attracted the attention of the
phyisists
 and mathematicians [8,12-14]. Needless to say, the mathematical structure of
  the QGS is very beautiful, however, to the physicist, they are more
interested
   in its application in the physics, In this respect, P. B. Wiegmann et al.
[16] and
    Y. Hatsugai et al. [17] has completed their oringinal exploration.
Certainly, it is still
     very interesting  to look for more applications of the QGS
    in  physics.

 In this letter, we will show that the QGS also can be found
 even in the simplest system of quantum mechanics and it is the origin of the
degeneracy of energy levels in our problem. In  more detail, with the help of
the
 representation theory of quantum group, we determined the degree of the
degeneracy
   of  Landau energy levels for such a system  which a spinless particle moves
    in a plane and experiences  a  uniform external magnetic field $\vec{B}$.



     We  consider a spinless particle which moves in a plane and
experiences an uniform external magnetic field along $z$-direction, $
\vec{B}=B\hat{e}_{z}$. The Hamiltanian of  system can be written
as
\begin{equation}
 H=\frac{1}{2m}(\vec{p}+e\vec{A})^2
\end{equation}
where $m$,$e$ are the mass and charge of  particle, respectively. $
\vec{A} $ is  vector potential which satisfy
\begin{equation}
\bigtriangledown \times \vec{A}=B\vec{e}_{z}
\end{equation}

The above problem can be easily solved in a proper gauge [1,3].
 In present paper, we study the gauge-independent case
with a periodical boundary condition(PBC).
 It is well known that in this  system there does not  exist the translational
invariance,
    however, it can exhibits
magnetic translation invariance which generated by the magnetic translation
operator [4] defined by
\begin{equation}
t(\vec{a})=\exp[\frac{i}{\hbar}\vec{a}\cdot(\vec{p}+
e\vec{A}+e\vec{r}\times\vec{B})]
\end{equation}
where $\vec{a}=a_{x}\hat{e}_{x}+a_{y}\hat{e}_{y}$ is an arbitrary
two-dimentional
vector.
The magnetic translation operator $t(\vec{a})$ satisfies the following group
property [5,6]:
\begin{equation}
t(\vec{a})t(\vec{b})=\exp[-i\frac{\hat{e}_{z}\cdot(\vec{a}\times\vec{b})}
{a_{0}^2}] t(\vec{b})t(\vec{a})
\end{equation}
where $a_{0}\equiv\sqrt{\frac{\hbar}{eB}}$ is the magnetic length.

Let
\begin{equation}
\vec{\kappa}=\vec{p}+e\vec{A}+e\vec{r}\times\vec{B}
\end{equation}

It is easy to prove that
\begin{equation}
 [t(\vec{a}),H]=0,\hspace{2 cm} [\vec{\kappa},H]=0
\end{equation}
which means that the system under consideration is invariant under
the magnetic translation transformation Eq.(3), and $\vec{\kappa}$ is
a conservative quantity.

With the help of the magnetic translation operator, one can construct the
following operators [7]:
\begin{eqnarray}
 J_{+}&=&\frac{1}{q-q^{-1}}[t(\vec{a})+t(\vec{b})],\nonumber  \\
 J_{-}&=&\frac{-1}{q-q^{-1}}[t(-\vec{a})+t(-\vec{b})],\nonumber  \\
 q^{2J_{3}}&=&t(\vec{b}-\vec{a}),\nonumber  \\
 q^{-2J_{3}}&=&t(\vec{a}-\vec{b})
\end{eqnarray}
with
\begin{equation}
 q=\exp(i2\pi\frac{\Phi}{\Phi_{0}})
\end{equation}
where $ \Phi=\frac{1}{2}\vec{B}\cdot(\vec{a}\times\vec{b}) $ is  magnetic
flux through the triangle  enclosed by vector $\vec{a}$ and $\vec{b}$,
$\Phi_{0}=\frac{h}{e}$ is  magnetic flux quanta.
A straitforward calculation shows that these operators $J_{+}$,$J_{-}$ and
$J_{3}$
satisfy the algebraic relation of the quantum group $sl_{q}(2)$ [8] as follows:
\begin{eqnarray}
 [J_{+},J_{-}]&=&[2J_{3}]_{q}   \nonumber    \\
 q^{J_{3}}J_{\pm}q^{-J_{3}}&=&q^{\pm 1}J_{\pm}
\end{eqnarray}
where we have used the following notation:
\begin{equation}
 [x]_{q}=\frac{q^{x}-q^{-x}}{q-q^{-1}}
\end{equation}
{}From Eqs.(6) and (7) it follows that:
\begin{equation}
 [J_{\pm},H]=0,\hspace{2 cm} [q^{\pm J_{3}},H]=0
\end{equation}
which indicates that $J_{\pm}$ and $J_{3}$ are conservative quantities of the
system. Therefore, there is the quantum group $sl_{q}(2)$ in the Landau problem
under our consideration.


Let $\Psi$  be wave function of  system in Schr\"{o}dinger picture.
In order to calculate explicitly the degree of degeneracy of Landau levels,
we impose the following PBC on the wave funtion[9]:
\begin{equation}
 t(\vec{L}_{1})\Psi =\Psi,\hspace{2 cm}  t(\vec{L}_{2})\Psi=\Psi
\end{equation}
where $\vec{L}_{1}=L_{1}\hat{e}_{x}$, and $\vec{L}_{2}=L_{2}\hat{e}_{y}$. This
boundary condition means that the particle is confined in a rectangular area of
 size $L_{1}\times L_{2}$.
{}From Eq.(12) it follows that the operators $t(\vec{L}_{1})$ and
$t(\vec{L}_{2})$
commute with each other. That is,
\begin{equation}
 t(\vec{L}_{1})t(\vec{L}_{2})=t(\vec{L}_{2})t(\vec{L}_{1})
\end{equation}
however, from Eq.(4) we have
\begin{equation}
t(\vec{L}_{1})t(\vec{L}_{2})=exp[-i\frac{\vec{e}_{z}\cdot(\vec{L}_{1}
\times\vec{L}_{2})}{a_{0}^{2}}]t(\vec{L}_{2})t(\vec{L_{1}})
\end{equation}
Combining Eq.(13) with Eq.(14) yields that
\begin{equation}
\exp(i2\pi\frac{\Phi}{\Phi_{0}})=1
\end{equation}
where $\Phi=\frac{1}{2}BL_{1}L_{2}$ is the magnetic flux through the triangle
enclosed
by $\vec{L}_{1}$ and $\vec{L}_{2}$.
 Eq.(15) implies that
\begin{equation}
\Phi=N_{s}\Phi_{0}
\end{equation}
where $N_{s}$ is a positive integer.
Therefore, the periodic boundary condition Eq.(12) is equivalent to  the
magnetic flux quantinization.

Notice that not all the translation operators $t(\vec{a})$ can keep the
boundary
condition Eq.(12) invariant. In other words,
\begin{equation}
 t(\vec{L}_{i})t(\vec{a})\Psi=t(\vec{a})\Psi   \nonumber \\
( i=1,2 )
\end{equation}
 can not be satisfied by an arbitrary maganetic translation $t(\vec{a})$.
However, if we define two primitive magnetic translation operators in the
following way [9]:
\begin{equation}
 T_{x}\equiv t(\frac{\vec{L}_{1}}{N_{s}}),\hspace{2 cm} T_{y}\equiv
t(\frac{\vec{L}_{2}}{N_{s}})
\end{equation}
One can find that only $T_{x},T_{y}$ and their integer powers can make Eq.(17)
hold.

By a straightforward calculation, it can be checked that the following
relations hold
\begin{eqnarray}
T_{y}T_{x}&=&exp(i\frac{2\pi}{N_{s}})T_{x}T_{y},\nonumber \\
 T_{y}T_{-x}&=&exp(-i\frac{2\pi}{N_{s}})T_{-x}T_{y}
\end{eqnarray}
\begin{eqnarray}
 T_{-y}T_{x}&=&exp(-i\frac{2\pi}{N_{s}})T_{x}T_{-y},\nonumber \\
 T_{-y}T_{-x}&=&exp(i\frac{2\pi}{N_{s}})T_{-x}T_{-y}
\end{eqnarray}
\begin{equation}
 T_{-x}T_{x}=T_{-y}T_{y}=1
\end{equation}
Making use of the operators $T_{\pm{x}}, T_{\pm{y}}$ and the above commutation
relations, we can construct a basic quantum group with the generators as
follows:
\begin{equation}
 J_{+}=\frac{-i}{q-q^{-1}}(T_{-x}+T_{-y}),\hspace{2
cm}J_{-}=\frac{-i}{q-q^{-1}}(T_{x}+T_{y})
\end{equation}
\begin{equation}
  K^{+2}=qT_{-y}T_{x},\hspace{2 cm}  K^{-2}=q^{-1}T_{-x}T_{y}
\end{equation}
where the deformation paramerter is given by
\begin{equation}
 q=\exp(i\frac{\pi}{N_{s}})
\end{equation}
It is easy to check that these generators obey the standard commutation
relations
of the quantum group $sl_{q}(2)$ [8]:
\begin{equation}
 [J_{+},J_{-}]=\frac{K^{+2}-K^{-2}}{q-q^{-1}},\hspace{2 cm}
K^{+}J_{\pm}K^{-}=q^{\pm 1}J_{\pm}
\end{equation}
We can also find that the generators $J_{\pm}$ and $K^{\pm}$ are conservative
quantities of the system under our consideration, namely
\begin{equation}
 [J_{\pm},H]=0,\hspace{2 cm} [K^{\pm },H]=0
\end{equation}
The above analysis indicates that $sl_{q}(2)$ is the basic symmetry in our
system.
Furthermore, according to the fundamental  principle of quantum mechanics,
Eq.(25) and Eq.(26)
 imply that there is degeneracy of  Landau levels in the system.

  In what follows we will discuss
 the relation between degeneracy of Landau levels and the cyclic representation
 of $sl_{q}(2)$.
 Since $N_{s}$ is an integer, from Eq.(24) we see that
 \begin{equation}
  q^{2N_{s}}=1
 \end{equation}
     which means that $q$ is a root of unity. In this case, the representation
of quantum
 group     has
 many exotic properties [10,11]. Typically, it has the  cyclic representation,
  which implies that there is neither highest weight nor the lowest weight
[10,11],
  and the dimension of the irreducible representation is $2N_{s}$ in the case
under
 our  consideration.

Furthermore, without loss of  generality, according to Eq.(26) we can
simutaneously
 diagonalize $H$ and $K^{\pm}$. In other words, one can choose a set of
  basis vectors $\left | n,k\right\rangle=\left |n\right\rangle\otimes\left
|k\right\rangle $
   to be the simutaneous eigenvectors of operators  $H$ and $K^{\pm}$.
That is, we can take
\begin{equation}
 H\left |n,k\right\rangle=E_{n}\left|n,k\right\rangle
\end{equation}
and
\begin{equation}
  K^{\pm}\left |n,k\right\rangle=q^{\pm (\lambda-2k-2\mu)}\left
|n,k\right\rangle
\end{equation}
where $n=0,1,...,\infty$ is the symbols of the energy level, and
 $k=0,1,...,2N_{s}-1$ is the new quantum numbers which distinguish the
 different quantum states in the same degenerate energy level.

  According to the representation theory of quantum group at root of unity
[12,13,14],
the actions  of the $sl_{q}(2)$ generators on these basis vectors are given
    by
 \begin{equation}
  J_{+}\left |n,k\right\rangle=[\lambda-\mu-k+1]\left | n,k-1 \right\rangle,
\hspace{1.2cm} (1\leq k\leq 2N_{s}-1) \nonumber
\end{equation}
\begin{equation}
  J_{+}\left |n,0\right\rangle=\xi^{-1}[\lambda-\mu+1]\left |
n,2N_{s}-1\right\rangle   \nonumber
\end{equation}
\begin{equation}
  J_{-}\left |n,k\right\rangle=\left | n,k+1 \right\rangle,\hspace{1.2cm}(0\leq
k \leq 2N_{s}-2)  \nonumber
\end{equation}
\begin{equation}
  J_{-}\left |n,2N_{s}-1\right\rangle=\xi\left | n,0 \right\rangle
\end{equation}
    where $\lambda,\xi,\mu $ are  constants determined by the cyclic properties
of the representation of  quantum group and the notation
$[x]=\frac{q^{x}-q^{-x}}{q-q^{-1}}$
 has been used.

    Since the dimension of the irreducible represention space ${\left
|n,k\right\rangle }$
is $2N_{s}$, from Eqs.(28), (29) and (30) we can see that the degree of
degeneracy of Landau levels
is just $2N_{s}$[15]. This is one of the main conclusions of this paper.
    In particular, when the boundary of the system approaches to the infinity
(i.e.
$L_{1}\rightarrow \infty,L_{2}\rightarrow \infty)$, we can see that
$2N_{s}\rightarrow \infty$. In this case, the system exihibits the continous
degeneracy, this is a well known result.

In summary,  we have shown that there is quantum group symmetry in the Landau
problem,
and the existence of the quantum group symmetry is independent of the choice of
the
gauge, and the degeneracy of  Landau levels in the system under our
consideration
originates from the quantum group symmetry $sl_{q}(2)$. We have found that
under
the PBC, the  degree of the degeneracy of Landau levels
is finite, and it is just the dimension of the irreducible cyclic
representation
of the quantum group $sl_{q}(2)$. When the boundary approaches to infinity, the
usual
result on the degeneracy of Landau level can be recovered.
It is worth mentioning that for a particle with spin, for instance,
an  electron moving in a plane, although  each energy level will split into two
due to the additional
Zeeman's energy, However, the degree of the degeneracy of  its energy levels
still keep $2N_{s}$
 except the ground state for which the  degree of the
 degeneracy is $N_{s}$. The reason is that  energy levels
  overlap between the upper level and lower level except the lowest one,
   which is just  the ground state.

\vspace{0.5cm}
\begin{flushleft} \Large \bf
Acknowledgments
\end{flushleft}
The authors acknowledge Dr. D. F. Wang for valuable discussions and useful
suggestions.
 This research is partly supported by the National Natural Science
Foundation of China.

\end{document}